\documentclass{article}
\usepackage{spconf,amsmath,graphicx}
\usepackage{amsthm,amssymb,bm,commath}
\usepackage{graphicx,tikz}
\usepackage{algorithmic,algorithm}
\usepackage{verbatim} % Package to include the {comment} environment to block out whole blocks of text easily

\DeclareMathOperator{\diag}{diag}

\title{Learning spatially-correlated temporal dictionaries \\for calcium imaging}
\twoauthors
{Gal Mishne\sthanks{ G.M. is supported by the NIBIB and the NINDS, of the NIH, Award R01EB026936.}}
	{Yale University\\
	Department of Mathematics\\
	New Haven, CT 06520 USA}
  {Adam S. Charles}
	{Princeton University\\
	Princeton Neuroscience Institute\\
	Princeton, NJ 08540 USA}
\begin{document}
\ninept
\maketitle
\begin{abstract}
Calcium imaging has become a fundamental neural imaging technique, aiming to recover the individual activity of hundreds of neurons in a cortical region. Current methods (mostly matrix factorization) are aimed at detecting neurons in the field-of-view and then inferring the corresponding time-traces. In this paper, we reverse the modeling and instead aim to minimize the spatial inference, while focusing on finding the set of temporal traces present in the data. We reframe the problem in a dictionary learning setting, where the dictionary contains the time-traces and the sparse coefficient are spatial maps.
We adapt dictionary learning to calcium imaging by introducing constraints on the norms and correlations of the time-traces, and incorporating a hierarchical spatial filtering model that correlates the time-trace usage over the field-of-view. 
We demonstrate on synthetic and real data that our solution has advantages regarding initialization, implicitly inferring number of neurons and simultaneously detecting different neuronal types.  
\end{abstract}
\section{Introduction}
\label{sec:intro}
Calcium imaging, or the optical recording of calcium concentrations in neural tissue, is an important neural imaging technique due to its ability to simultaneously record large neural populations at single cell resolution in awake behaving animals~\cite{RN2,RN25,RN27}. 
This technique enables neuroscientists to investigate the role of large cortical areas, uncovering neural representations that are easily lost in single-unit recordings.
In particular, calcium imaging via two-photon microscopy (TPM) leverages nonlinear optical interactions to resolve activity at greater depths, enabling minimally-invasive imaging hundreds of microns beneath the surface. 
The advances in scan rate, achievable field-of-view (FOV), and fluorescence signal strength that have solidified TPM as a fundamental technique in neuroscience, however, have also created a complex data-analysis problem. The high-dimensionality of the datasets ($10$K-$100$K $512 \times512$ frames) and the complex noise environments must be overcome to accurately extract the individual neurons' time-traces. Volumetric methods that project 3D volumes to 2D images further complicate demixing by producing highly overlapping neuronal structures~\cite{song2017volumetric}. 

Current TPM analysis methods focus on factoring the data cube into a set of spatial profiles\footnote{Spatial profiles are sometimes termed Regions of Interest (ROIs), however we prefer the terminology ``profile'' as it more accurately reflects that these components are 2D projections of neural processes.}
(the location within the FOV that a neuron occupies) and a corresponding set of time-traces (the fluorescence activity for each component). 
While many methods have been devised to isolate these factors, based on deep learning~\cite{apthorpe2016automatic}, active contours~\cite{Reynolds2017}, spectral embeddings~\cite{Mishne2018}, matrix factorization remains the most popular approach~\cite{mukamel2009automated, maruyama2014detecting,pnevmatikakis2013sparse, pachitariu2016suite2p, pnevmatikakis2016simultaneous}. 
Given an $N_x \times N_y$ pixel FOV sampled at $T$ time-points, we have
\begin{equation}
\label{eq:factor}
\bm{Y}^T = \bm{A}\bm{\Phi}^T + \bm{E}^T,
\end{equation}
where $\bm{Y}\in\mathbb{R}^{T \times N_x N_y}$ is the fluorescence movie data, $\bm{A}\in\mathbb{R}^{N_x Ny \times M}$ are the spatial profiles for the $M$ neural components, $\bm{\Phi}\in\mathbb{R}^{T\times M}$ are their corresponding time-traces and $\bm{E}\in\mathbb{R}^{T \times N_x N_y}$ is the recording noise, often modeled as $i.i.d.$ Gaussian. 
Constraints are often added to this model to better represent the data, such as sparse compact spatial support, non-negativity and auto-regressive time-trace models~\cite{mukamel2009automated,pnevmatikakis2016simultaneous,SCALPEL}.
This matrix factorization equally weights both the spatial and temporal components. Ensuing validation likewise often assesses the accuracy of spatial profiles~\cite{berens2017standardizing}.

\begin{figure*}[ht!]
        \centering
        \includegraphics[scale = 0.95]{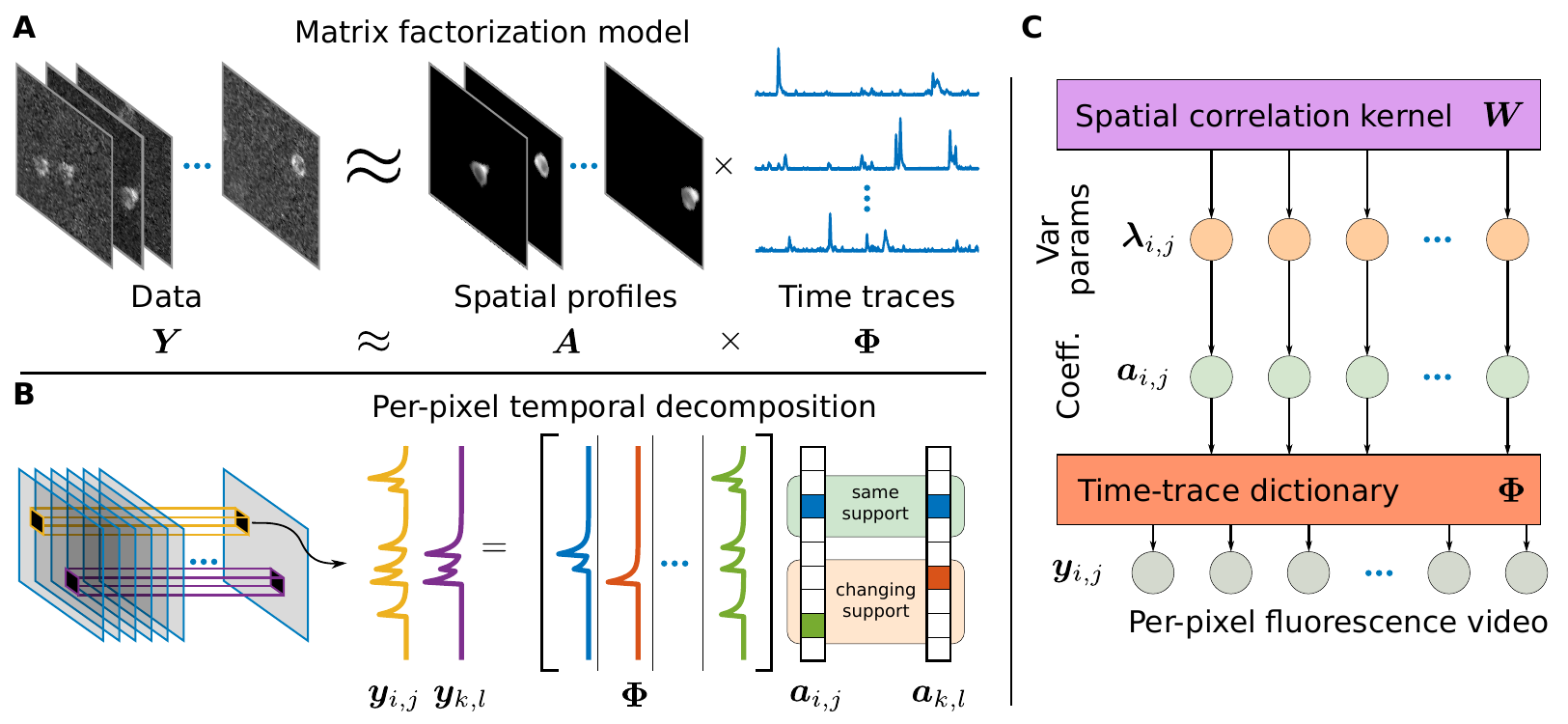}
\caption{A: Matrix factorization decomposes a fluorescence video into spatial profiles and time-traces. B: Our model emphasizes the role of time-traces and focuses on pixel-wise decompositions. C: Graphical model of spatial-filtered sparse coding that allows us to use less restrictive spatial information and introduce less biases into the time-trace estimates.}
     \label{fig:motivation}
     \vspace{-0.2in}
\end{figure*}

The main object of interest, however, for scientific inquiry is the set of time-traces, as they are the variables that are linked to behavior, stimuli, learning, etc. 
Thus, in this work, we 
reverse the modeling philosophy of looking for spatial components in order to find temporal traces, and rather directly model temporal traces.
Mathematically we transpose the original model as
\begin{equation}
\label{eq:factor2}
\bm{Y} = \bm{\Phi}\bm{A}^T + \bm{E},
\end{equation}
such that we reverse the roles of $\bm{\Phi}$ and $\bm{A}$. While a seemingly trivial semantic change, we will show this is a more natural model, and this paradigm reduces sensitivity to initializations and removes the burden of morphological post-processing. 
Philosophically this reorganization places more importance in finding the temporal activity and treats the spatial statistics as secondary, as opposed to typical analyses~\cite{pachitariu2013extracting,Diego2014,SCALPEL}.
This modeling shift is similar to the difference, for example, in modeling the spatial statistics of hyperspectral imaging data versus the predominant spectral end-member analysis.

To infer $\bm{\Phi}$ under the new model, we adopt a dictionary learning (DL) approach, where the dictionary is composed of temporal components and the sparse coefficients are the spatial profiles (Fig.~\ref{fig:motivation}B,C). %~(Sec. \ref{sec:dictionary}). 
In the temporal domain, we introduce correlation- and continuity-based regularizers on the learned dictionary. 
This enables implicit inference of the number of neurons, which remains a challenge for many current methods.
We further induce implicit spatial contiguity into the DL framework via Reweighted-$l_1$ Spatial Filtering (RWL1-SF) in the coefficient inference~\cite{charles2014spectral}.
We validate our spatially-filtered DL algorithm on both synthetic and real TPM data.

\section{Background}

Dictionary learning (DL) is an unsupervised method aimed at finding optimal, parsimonious representations for data given exemplar data~\cite{olshausen1996emergence,aharon2006rm}. In particular, DL decomposes the dataset $\bm{Y}$ in Equation~\eqref{eq:factor2} into the dictionary $\bm{\Phi}$ and coefficients $\bm{A}$, under the assumption that the rows of $\bm{A}$ are \emph{sparse} (i.e., most elements of $\bm{A}$ are zero). %The learned dictionary $\bm{\Phi}$  a parsimonious representation of $\bm{Y}$. 
While originally used for image processing tasks such as denoising or inpainting~\cite{elad2010role}, DL has been successful in other domains as well. 

Hyperspectral Imagery (HSI) has been one such application~\cite{charles2011learning,zhou2012nonparametric} that has many parallels to TPM. 
Both are high-dimensional image volumes, where the non-spatial dimension contains important identity information (time-courses in TPM holding neural identity and, optical reflection spectra in HSI holding material identity). 
Thus HSI processing has focused in large part on learning spectral signatures, using DL for unsupervised extraction of material spectra from data-cubes~\cite{charles2011learning}. 
In this case the dictionary contains spectral atoms and the sparse coefficient are spatial abundance profiles.  
Related to the concept of endmembers (the boundaries of the convex hull containing the HSI data~\cite{greer2012sparse}), the spectral dictionary both uncovers valuable semantic material information about the imaged area, and is used for many signal processing tasks (e.g. inpainting, super-resolution~\cite{xing2012dictionary,charles2014spectral}). 
To leverage the coarser spatial information, advanced models correlate the pixel-decompositions of neighboring pixels, for example using joint spatial-spectral dictionaries~\cite{xing2012dictionary}, coefficient regularization~\cite{chen2011sparse}, or spatial filtering~\cite{charles2014spectral}. 

Current applications of DL for TPM data, however, have focused on learning spatial dictionary atoms~\cite{pachitariu2013extracting,Diego2014,SCALPEL}.
For example Pachitariu et al.~\cite{pachitariu2013extracting} developed a spatial generative model based on convolutional sparse block coding with the goal of learning the spatial locations of somas and dendrites from the mean image across time. 
Diego and Hamprecht~\cite{Diego2014} extend convolutional sparse coding to video data, extracting the spatial components and their temporal activity while estimating a non-uniform and temporally varying background. 
SCALPEL~\cite{SCALPEL} similarly build a dictionary of spatial components that is refined through iterative merging and clustering, before the temporal components are inferred.

\begin{figure*}[ht!]
        \centering
        \includegraphics[scale=0.99]{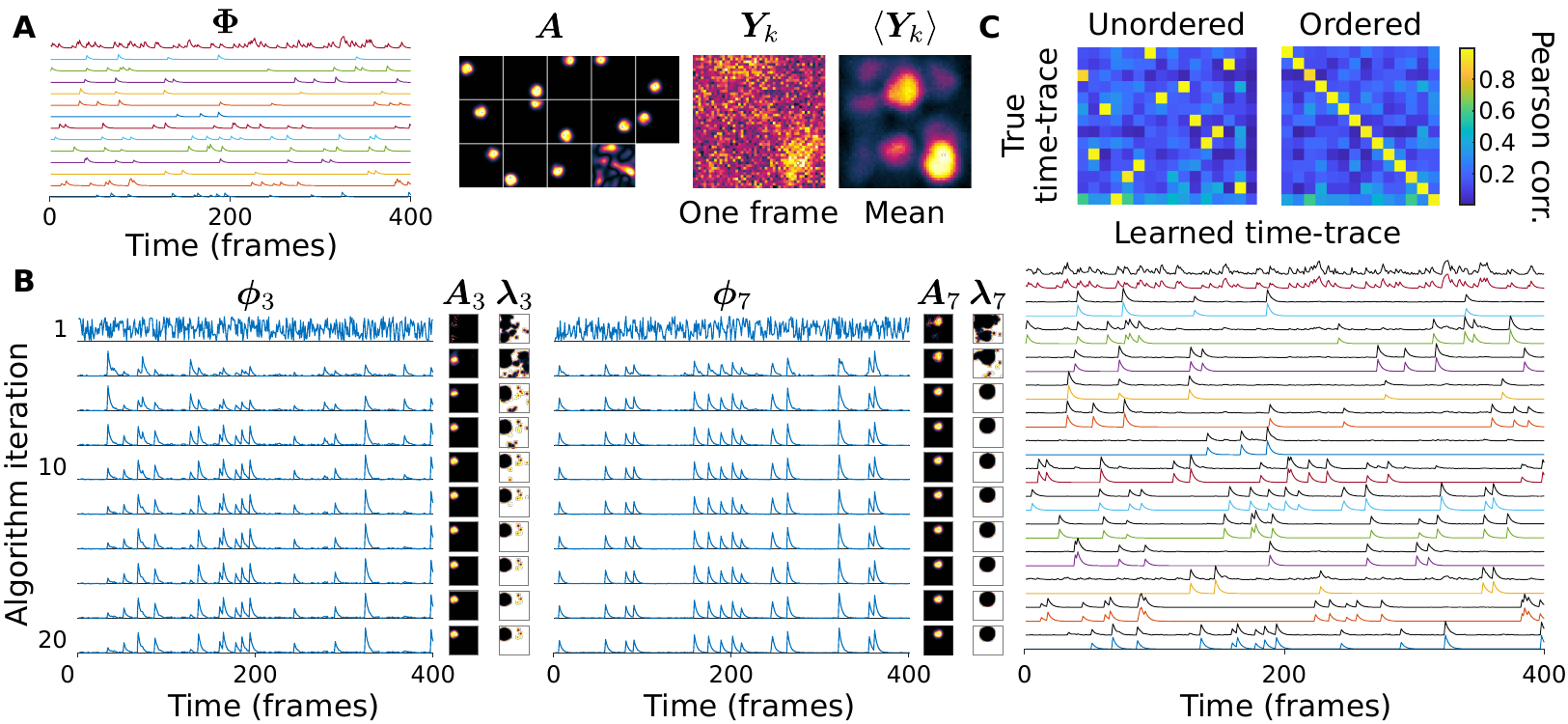}
        \caption{A: Simple simulation creates neural time courses ($\bm{\Phi}$) and mixes contiguous pixels ($\bm{A}$) together to simulate a fluorescence movie ($\bm{Y}$). B: Two example learned time traces adapting throughout the learning procedure. The iterative nature of our algorithm begins with random time-courses and converges on a local minimum of the factorization problem. C: Comparison of time traces to the true underlying time-traces demonstrate that the true underlying time-courses are recovered both for localized and distributed components (e.g., neurons and neuropil). }
     \label{fig:simulations}
     \vspace{-0.2in}
\end{figure*}

\section{Dictionary Learning for Calcium Imaging}
\label{sec:dictionary}

Following the inspiration from HSI data processing, we focus on modeling TPM data in terms of the time-traces of single neurons.
In particular, we combine DL methodology with the spatially correlated generative model of~\cite{charles2014spectral}.
This combination enables us to focus on accurate time-trace estimation, reducing errors that may arise when also trying to accurately estimate the spatial maps (i.e., sensitivity to initialization, errors due to mismatched spatial profiles, etc.).

Given a sequence of motion-corrected TPM frames $\bm{Y}_l\in\mathbb{R}^{N_x\times N_y}$, we consider the temporal activity vector $\bm{y}_{i,j}\in\mathbb{R}^T$ at each pixel $\{i,j\}$. 
In any given video, we model $\bm{y}_{i,j}$ as
\begin{gather}
    \bm{y}_{i,j} = \sum_{k} \bm{\phi}_k a_{i,j,k} + \bm{\epsilon}_{i,j},
\end{gather}
where $\bm{\phi}_k\in\mathbb{R}^T$ for $k\in[1,\cdots,K]$ are the $K$ neural time-traces, $a_{i,j,k}$ is the strength of each neuron's fluorescence at each pixel, and $\bm{\epsilon}_{i,j}\in\mathbb{R}^T$ represents the sensor noise. We define a cost function of the time-traces $\bm{\phi}_k$ and the spatial presence coefficients $a_{i,j,k}$, that captures both the data model and the \emph{a-priori} information that few neurons overlap at any given pixel ($\bm{a}_{i,j}=[a_{i,j,1},\dots,a_{i,j,K}]^T\in\mathbb{R}^K$ is sparse) and that both $\bm{\Phi}$ and $\bm{A}$ are non-negative. Mathematically, the basic DL problem is
\begin{gather}
\label{eq:DL}
      \arg\min_{\bm{\Phi},a_{i,j}} \left[ \sum_{i,j} \left\| \bm{y}_{i,j} - \bm{\Phi}\bm{a}_{i,j}\right\|_2^2 + \lambda\left\|\bm{a}_{i,j}\right\|_1 \right],
\end{gather}
where $\bm{\Phi} = [\bm{\phi}_1,\dots, \bm{\phi}_K]\in\mathbb{R}^{T\times K}$ is the time-trace dictionary, and $\lambda$ is a parameter that trades off data fidelity and sparsity in $\bm{a}_{i,j}$. 

Standard methods for solving Equation~\eqref{eq:DL} alternate between optimizing $\bm{a}_{i,j}$ given an initial $\bm{\Phi}$, which is in turn updated based on the inferred $\bm{a}_{i,j}$. As there is no inherent penalization on $\bm{\Phi}$, a constraint on the norm or maximum value of each $\bm{\phi}_k$ is usually imposed to prevent the trivial solution of $\|\bm{\Phi}\|_F^2 \rightarrow \infty$. 
We extend this basic model for TPM data in two important ways. First build in additional constraints on $\bm{\Phi}$ to isolate single traces despite modeling nonlinear data with a linear-generative model.    
Second we expand the sparsity model over $\bm{a}_{i,j}$ to include spatial contiguity.

\noindent \textbf{Temporal dictionary learning:} 
We introduce two penalties over $\bm{\Phi}$ to the cost function to facilitate both automatically adapting the number of neural time-traces, and reducing the impact of subtle nonlinearities in the spatial expression of a given time-trace.
A Frobenius cost $\|\bm{\Phi}\|_2^2$ serves to remove unused dictionary elements, and a penalty over intra-dictionary correlations $\|\bm{\Phi}^T\bm{\Phi} - \diag(\bm{\Phi}^T\bm{\Phi} )\|_{sav} = \sum_{i\neq k}\bm{\phi}_i^T\bm{\phi}_k$ penalizes time-traces with trivial differences. 
To ensure stable convergence, we also include a continuation term that penalizes the change in $\widehat{\bm{\Phi}}$ from the previous estimate $\widetilde{\bm{\Phi}}$. Overall, the update for $\bm{\Phi}$ given $\bm{a}_{i,j,k}$ is
\begin{gather}
\begin{split}
        \widehat{\bm{\Phi}} & = \arg\min_{\bm{\Phi}\geq 0} \left\|\bm{Y} - \bm{\Phi}\bm{A} \right\|_F^2 + \kappa_1\|\bm{\Phi}\|_F^2 \\ &+  \kappa_2 \|\bm{\Phi} - \widetilde{\bm{\Phi}}\|_F^2 + \kappa_3 ||\Phi^T\Phi-\textrm{diag}(\Phi^T\Phi)||_{sav} \label{eq:PhiUpdate}
\end{split}
\end{gather}

\noindent \textbf{Spatially filtered sparse coding:}
To remedy the lack of spatial cohesion in traditional DL, we adapt the re-weighted $\ell_1$ spatial filtering (RWL1-SF) previously used to infer sparse, clustered coefficients~\cite{charles2014spectral} (Fig.~\ref{fig:motivation}C). 
RWL1-SF, which is an expansion of re-weighted $\ell_1$~\cite{candes2008enhancing} and the Laplacian-scale mixture model~\cite{GAR:2010}, places a hierarchical layer above the Laplacian prior (Fig.~\ref{fig:motivation}C). Specifically the data $\bm{Y}$ at each pixel $\{i,j\}$ given $\bm{\Phi}$ is modeled as
\begin{eqnarray}
        \bm{y}_{i,j}    & \sim & \mathcal{N}\left( \bm{\Phi} \bm{a}_{i,j}, \sigma_y^2 \bm{I}\right) \\
        a_{i,j,k}       & \sim & \mbox{Lap}\left( [\bm{W}\ast\lambda_k]_{i,j} \right)                                    \\
        \lambda_{i,j,k} & \sim & \mbox{Gamma}(\alpha, \theta)                                                            \\
        \bm{\phi}_k     & \sim & \mathcal{N}\left(\bm{0}, \bm{\sigma}_{\phi}^2\bm{I} \right)   
\end{eqnarray}

Inferring $\bm{a}_{i,j}$ (via approximate expectation-maximization) under this model amounts to iteratively solving
\begin{gather}
        \widehat{\bm{a}}_{i,j} = \arg\min_{\bm{a}\geq0} \frac{1}{2\sigma_y^2}\left\|\bm{y}_{i,j} - \bm{\Phi}\bm{a} \right\|_2^2 + \sum_{k} \lambda_{i,j,k}|a_{i,j,k}|,  \label{eq:AUpdate}
\end{gather}
for all $\{i,j\}$ and then updating the weights $\lambda_{i,j,k}$ as
\begin{gather}
\label{eq:weights}
        \lambda_{i,j,k} = \xi\left(\beta + |a_{i,j,k}| +  \left[|\bm{W}\ast\widehat{\bm{A}}_{k}|\right]_{i,j}\right)^{-1}, 
\end{gather}
where $\bm{A}_k\in\mathbb{R}^{N_x \times N_y}$ contains the presence coefficients for a single neuron across the FOV and $\bm{W}\ast\widehat{\bm{A}}_{k}$ indicates a 2D convolution. Here $\xi$ and $\beta$ depend only on model parameters $\alpha$, $\theta$, $\sigma_y^2$~\cite{charles2014spectral}.

The weights $\lambda$ in RWL1-SF incorporate spatial information into per-pixel solutions by sharing second-order statistics. 
Coefficients ``activated'' in the initial optimization lower the weights for neighboring coefficients, encouraging them to activate in subsequent iterations, whereas non-active coefficients penalize activation with higher weights. 
The kernel $\bm{W}$ specifies the coefficient influence radius, dictating the neighborhood where interactions are strongest.

\begin{algorithm}
    \caption{Stochastic Filtered CIDL}
    \label{alg:alg-curr}
    \begin{algorithmic}
        \STATE Input: data $\bm{Y}$, parameters $\{\beta$,$\xi$,$\kappa_1$,$\kappa_2$,$\kappa_3$,$\bm{W}\}$
        \WHILE{not converged}
            \FOR{all voxels}
                \STATE Initialize $\lambda_{i,j,k} = 1 \forall \{i,j,k\}$
                \FOR{$l \leq 3$}
                   \STATE Update $\widehat{\bm{a}}_{i,j}$ via Equation~\eqref{eq:AUpdate} 
                   \STATE Update $\lambda_{i,j,k}$ via Equation~\eqref{eq:weights} 
                \ENDFOR
            \ENDFOR
            \STATE Update $\widehat{\bm{\Phi}}$ via Equation~\eqref{eq:PhiUpdate} 
        \ENDWHILE
    \end{algorithmic}
\end{algorithm}

\begin{figure}
\includegraphics[width=0.5\textwidth]{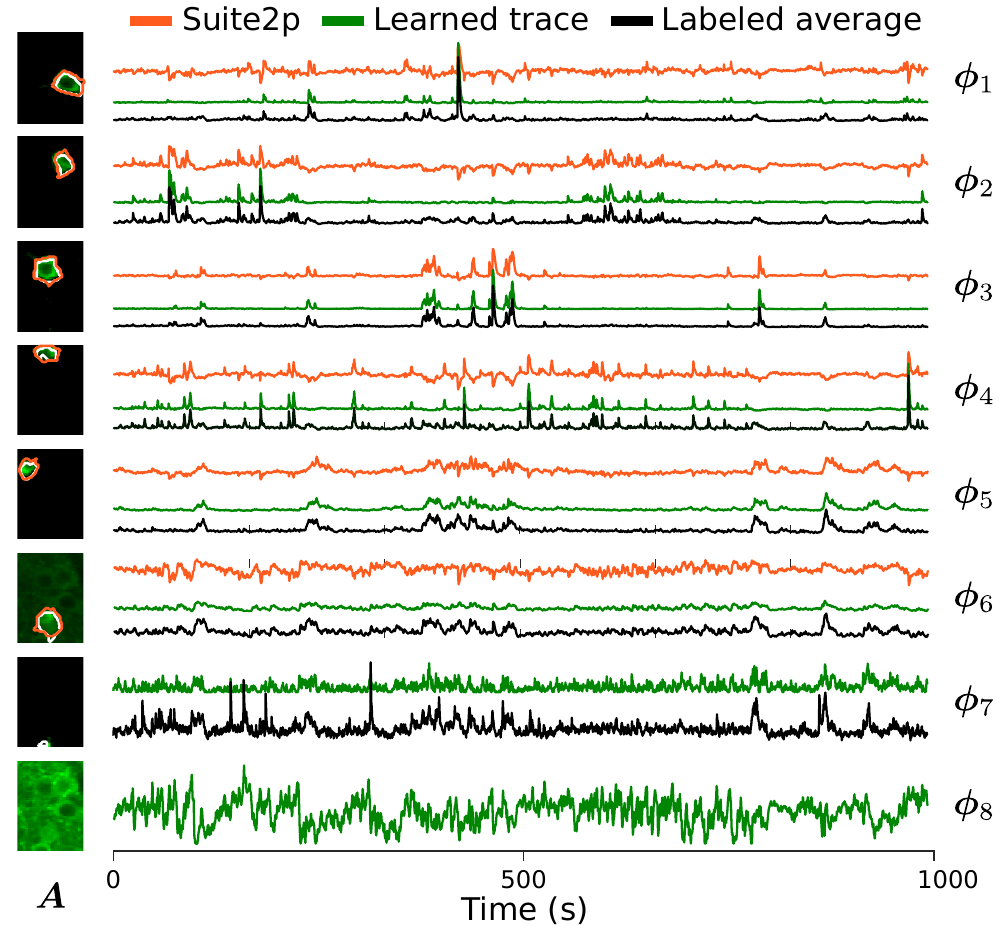} 
\caption{Learning temporal dictionaries for real calcium imaging data. The seven ground truth components (white outlines: black traces inferred via spatial projections), and one background component, were found with our method (green traces). Suite2p~\cite{pachitariu2016suite2p} (orange traces) found only 6 time-traces and with significant baseline fluctuation.}
     \label{fig:results1}
     \vspace{-0.2in}
\end{figure}

\section{Experimental Results}
\label{sec:results}

\noindent \textbf{Simulated data: }
We first validate our method on simulated data composed of 14 time-traces generated by convolving a randomly generated weighted spike train with an autoregressive filter with an single pole at -0.7 (Fig.~\ref{fig:simulations}A). Spatial presence maps were created via draws from a 2D Gaussian Process which was windowed by a truncated 2D Gaussian filter. One component was left unwindowed to simulate distributed neuropil signals. 

We learned time-traces for this data using Algorithm~\ref{alg:alg-curr} with 16 components, $\kappa_1=0.3$, $\kappa_2=0.4$, $\kappa_3=0.2$, $\xi=2$, and $\beta=0.1$. While the stopping criteria was $\|\Delta\widehat{\bm{\Phi}}\|_F^2/\|\widehat{\bm{\Phi}}\|_F^2 \leq 10^{-5}$, the algorithm completed in under 20 iterations. Most time-traces converged quickly to a gross estimate and only took longer to remove small, spurious transients (Fig.~\ref{fig:simulations}B). The learned dictionary correctly identified the 14 time-traces, and also correctly reduced the extra time-traces to negligible levels (Fig.~\ref{fig:simulations}C). These results encouraged us to apply this method to real two-photon calcium imaging data. % $\lambda_0=10$, 

\noindent \textbf{Somatic Imaging}
We ran our approach on a $61 \times 46$~pixels by 3000~frames (at 3Hz) dataset from Neurofinder~\cite{berens2017standardizing}. 
We initialize the dictionary with 14 components, as compared to the 7 labeled ground-truth neurons. 
We compare our results with Suite2p~\cite{pachitariu2016suite2p} which is a modern, state-of-the-art ROI extraction method.   

Seven of the resulting learned time-traces (Fig.~\ref{fig:results1}, green) display a close match with the time-traces estimated from the ground-truth labeled pixels (Fig.~\ref{fig:results1}, black) and contain less neuropil induced dips than the Suite2p estimates (Fig.~\ref{fig:results1}, orange). 
Of the 7 labeled neurons, we found 7 neurons and a background component while Suite2p found 6. 
Additionally, we detect 3 apical dendrites and a fourth component composed of basal dendrites not included in the provided ground-truth. Detecting these dendrites is possible because of our modeling emphasizing the temporal decompositions rather than relying on specific initialization of spatial components, or a-priori assumptions of neuron's sizes and shapes.  

\begin{figure}
\includegraphics[width=0.5\textwidth]{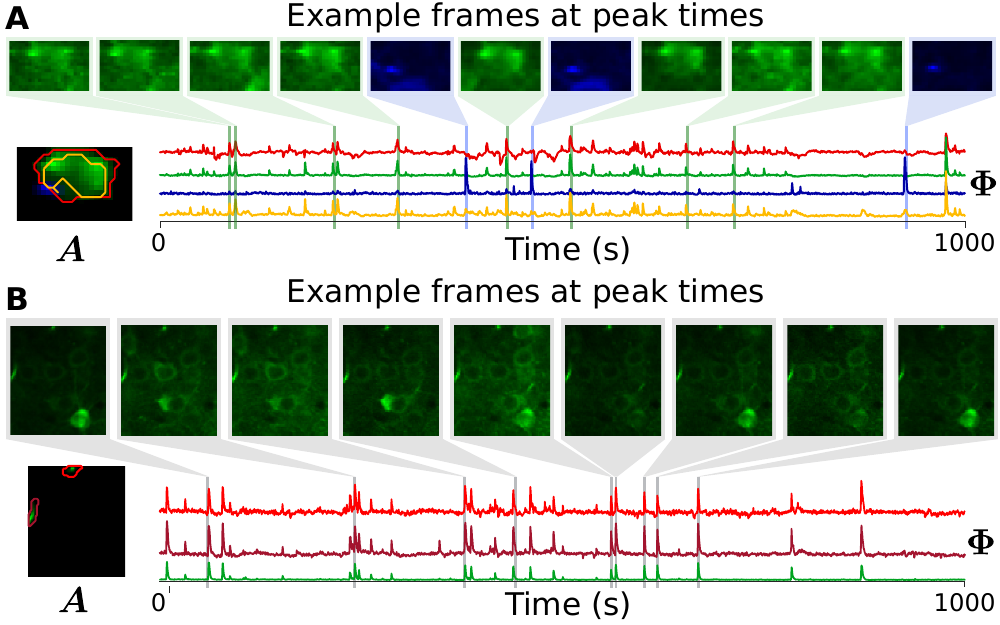} 
\caption{Temporal DL was able to find subtle features in the Neurofinder dataset. A: An apical dendrite (blue) significantly overlapping with a larger soma (green) was isolated. B: spatially disjoint pieces of the same neuron were correctly identified as one time-trace, while the spatial priors in Suite2p split the component into two.}
     \label{fig:results2}
     \vspace{-0.2in}
\end{figure}

One interesting case was an apical dendrite located adjacent to a soma (Fig.~\ref{fig:results2}A). 
Several example frames from the raw video at peaks of activity for one of the components shows that there is indeed an apical (blue) with separate activity from the soma (green).
Averaging the activity of the pixels in the ground-truth labeling (black trace) picks up the activity from the dendrite (blue trace).
Suite2p is also effected by this overlap, depicting dips in the baseline estimate every time the apical dendrite fires. 
A second interesting case from a different region of the same Neurofinder dataset (Fig~\ref{fig:results2}B) has our method inferring one dictionary element (green) that is active in two locations that Suite2p marks as two different components (red and burgundy).  
Example frames from the video confirm that these two components always fire together (Fig~\ref{fig:results2}B).

\section{Conclusions}
We propose a new approach for extracting neural time-traces from two-photon calcium imaging data. Our approach is based on expanding a dictionary learning algorithm that focuses on time-trace estimation by leveraging weaker spatial information than other methods that rely heavily on (e.g., morphological classification). We validate our method both in simulation and real calcium imaging data. Our method correctly picks out known components in both cases, and furthermore can uncover more subtle features in the real datasets, such as spatial distant pieces of the same neuron. 
Furthermore, the lack of overly-specific spatial regularization means that our method is applicable to other imaging modalities, such as wide-filed calcium imaging, where exact spatial structure is difficult to quantify.  

A number of practical considerations arise in our method, specifically initialization, selecting the number of neurons, and parameter selection.  
First, we initialize $\bm{\Phi}$ with random values, demonstrating a reduced sensitivity to initialization than other approaches~\cite{pnevmatikakis2016simultaneous}. It may still be possible to improver performance with targeted initializations, however; a possibility that should be explored in future work.
Second, the number of dictionary components should be set to more than the expected number of neurons and a background components. The sparsity and Frobenius norms serve to decay unused components (implicitly estimating the number of neurons), but cannot add new elements.    
Finally, we hand-tuned a number of parameters, e.g., setting the kernel $W$ as a $7 \times 7$ Gaussian kernel with variance 3. Future work should address automatic parameter selection (e.g., using variational methods) and sensitivity analysis.  

\newpage 
\clearpage
\bibliographystyle{IEEEbib}
%\bibliography{ci_dl.bib}

\end{document}